\documentclass[startpage={1}, year={25}]{fmcad}

\usepackage{graphicx} 
\usepackage{amsmath}  
\usepackage{amssymb}  
\usepackage{amsthm}   
\usepackage{booktabs} 
\usepackage{cite}     
\usepackage{cleveref}

\title{Two Strategies to Improve the St\aa lmarck Procedure}

\author{
    \IEEEauthorblockN{Sergei Leonov}
    \IEEEauthorblockA{Amherst College\\
    sleonov27@amherst.edu}
    \and
    \IEEEauthorblockN{Liam Davis}
    \IEEEauthorblockA{Amherst College\\
    ljdavis27@amherst.edu}
}

\begin{document}

\maketitle

\begin{abstract}
In this paper, we introduce StalmarckSAT, the a modern re-implementation of the St\aa lmarck Procedure for SAT solving, and present two novel strategies to improve the Procedure, Cardinality Driven Branching (CDB) and Deductive Priority Ordering (DPO). CDB is a heuristic to improve branching with the dilemma rule, and DPO intelligently orders simple rules based on their deductive potential. Our results demonstrate improved solve times with both strategies.
\end{abstract}

\section{Introduction}
SAT solving has impacted many fields like theorem proving, model checking, and verification. While modern CDCL-based SAT solvers have improved significantly in recent years, the St\aa lmarck Procedure, patented in 1989 and presented at FMCAD 1998, has remained almost untouched for 30 years. The procedure was originally touted for its applicability to industrial-scale problems, as the performance is sensitive to problem difficulty rather than problem size. This suggests that with more research, the St\aa lmarck Procedure may be suitable to certain classes of problems where CDCL struggles. Inspired to revamp this procedure, we introduce StalmarckSAT, a modern re-implementation of the St\aa lmarck Procedure along with two optimization strategies.

\section{Related Work}

The St\aa lmarck Procedure was introduced by St\aa lmarck in his original tutorial in 1998 \cite{stalmarck1998tutorial}. Since then, Bjork generalized the procedure to first order logic \cite{bjork2009first}, Cook explored using the procedure to prove unsatisfiability of boolean formulas containing inequalities \cite{cook2005using}, and Thakur analyzed the procedure in the context of abstract interpretation \cite{thakur2012generalization}. Despite its potential, the lack of an accessible, open source implementation has hindered research, which motivated us to create StalmarckSAT.

\section{Background}
In this section, we provide a quick introduction to the St\aa lmarck Proof System. A full formulation is available in St\aa lmarck's original tutorial \cite{stalmarck1998tutorial}.

\subsection{Normalized Form}
The St\aa lmarck Procedure operates on implications of propositional variables \cite{thakur2012generalization} \cite{cook2005using}. The transformation is completed in linear time with the following rules:

\begin{enumerate}
  \item $A \lor B \quad \Rightarrow \quad \neg A \rightarrow B$
  \item $A \land B \quad \Rightarrow \quad \neg (A \rightarrow \neg B)$
  \item $\neg \neg A \quad \Rightarrow \quad A$
  \item $\neg A \quad \Rightarrow \quad A \rightarrow \text{False}$
\end{enumerate}

The formula is then encoded into implication triplets, where a formula $x\leftrightarrow y \rightarrow z$ is represented as a tuple $(x,y,z)$. To convert normalized formula into set of triplets, intermediate bridge variables that represent the values of subexpressions are assigned. For example, the formula $(p \land \neg q) \lor r$ can be stored as a triplet pair 
$\{(b_1, q, r), (b_2, p, b_1)\}$. 

\subsection{Simple Deduction Rules}
The St\aa lmarck procedure starts with incomplete analysis with a set of simple rules derived from the implication truth table. The rules are as follows:

\[
\begin{array}{rlrlrl}
(1)\; & \cfrac{(0, y, z)}{\substack{y / 1 \quad z / 0}} 
& (2)\; & \cfrac{(x, y, 1)}{\substack{x / 1}} 
& (3)\; & \cfrac{(x, 0, z)}{\substack{x / 1}} \\[1.2ex]
(4)\; & \cfrac{(x, 1, z)}{\substack{x / z}} 
& (5)\; & \cfrac{(x, y, 0)}{\substack{x / \neg y}} 
& (6)\; & \cfrac{(x, x, z)}{\substack{x / 1 \quad z / 1}} \\[1.2ex]
(7)\; & \cfrac{(x, y, y)}{\substack{x / 1}} 
&       &  
&       &  
\end{array}
\]

\subsection{Branching with the Dilemma Rule}
For complete analysis, when no simple rules can be applied, the dilemma rule is used to branch on a chosen literal when no simple rules apply. For example, given a triplet pair $\{(1, \neg p, p), (1, p, \neg p)\}$, we can branch on the two polarities of literal $p$: when $p=0$, the triplets become $(1, 1, 0)$ and $(1, 0, 1)$ respectively; and when $p=1$, they become $(1, 0, 1)$ and $(1, 1, 0)$ respectively.

An application of the dilemma rule results in three cases. First, if both branches lead to a contradiction, there is no counterexample to the original formula, so the negation of the original formula is valid. Next, if only one branch leads to a contradiction, the non-contradictory branch is explored further using simple rules. Lastly, if neither branch leads to a contradiction, the current assignment is a counterexample. Therefore, the negation of the original formula is not valid.

In the example above, both branches contradict Rule 1 ($1 \neq 1 \rightarrow 0)$. Thus, there is no counterexample, and negated formula is a tautology.

\section{Methodology}

In this section, we detail our open source implementation, then present two strategies to improve the baseline St\aa lmarck Procedure, Cardinality Driven Branching (CDB) and Deductive Priority Ordering (DPO).

\subsection{StalmarckSAT}

For experimentation, we developed \href{https://github.com/Stalmarck-Satisfiability/StalmarckSAT}{StalmarckSAT}, a Rust-based implementation of the St\aa lmarck Procedure and the only public implementation on GitHub. Like modern SAT solvers, the solver operates on CNF formulas, and is designed to encourage research on the procedure. Presently, the solver integrates the baseline procedure as well as CDB and DPO. In its current form, the solver is not competitive with modern SAT solvers, but is a useful tool to conduct research.

\subsection{Cardinality Driven Branching for the Dilemma Rule}

The core principle of CDB is to prioritize branching on variables appearing with greater frequency (or higher cardinality) within the normalized triplet representation, mirroring VSIDS heuristics in CDCL solvers \cite{liang2015understandingvsidsbranchingheuristics}. The implementation of CDB is done in two stages, variable pre-processing and selection.

\subsubsection{Pre-processing}
CDB initiates a single pass over the entire collection of triplets, constructing a frequency map that keeps track of the frequencies of each variable. The frequency map is then converted into a sorted list of variables ordered in descending order of their cardinality, providing an immediate lookup for the highest cardinality variables. 

\subsubsection{Variable Selection for the Dilemma Rule}
When no simple rules are available to be applied, CDB searches for the original unassigned variable with the highest cardinality, excluding bridge variables. This prioritizes variables that are part of the initial problem, with the goal of resolving the most impactful constraints first. If all original variables have already been assigned, CDB selects the unassigned bridge variable with the highest cardinality. 

\subsection{Deductive Priority Ordering (DPO) of Simple Rules}

DPO is a heuristic strategy that intelligently selects the application order of simple rules. Its goal is to prioritize rules most likely to trigger further deductions. DPO assigns a deductive potential score to each triplet before the main solving loop.

\subsubsection{Triplet Scoring}
A triplet's score combines a base frequency score and deductive bonus. The base frequency score is the sum of the frequencies of its constituent variables, gathered during CDB's preprocessing step. The base score is then multiplied by a the deductive bonus that estimates the associated simple rule's deductive power. A triplet $(x, x, z)$ corresponds to a rule that deduces two variable assignments, and receives a bonus multiplier of 3. A triplet $(x, y, y)$ corresponds to a rule that deduces one variable assignment, and receives a bonus multiplier of 2. All other triplet structures receive no bonus.

\subsubsection{Pre-sorting and Application}
After the initial scoring, the entire list of triplets is sorted in descending order by DPO score, and rules are applied to the highest-scoring triplets first. With this, DPO ensures that the most promising simple rules are applied at each step.

\section{Experiments and Results}

\subsection{Experimental Setup}
We compared CDB and DPO to the baseline implementation of the St\aa lmarck Procedure on StalmarckSAT on 1000 KSAT benchmarks. Each KSAT had 50 variables, 218 clauses, and a clause length of 3. Experiments were run on a server with 2.6-GHz AMD CPUs with 4 CPU cores allocated per benchmark. Each benchmark was given a 30 minute time limit and a 32 GB memory limit.

\subsection{Results}

Table \ref{tab:table} reports the number of instances solved within the time limit and average solve time on solved instances for each configuration. Overall, both DPO and CDB led to a substantial increase in solved instances, and combining both strategies led to the most solved instances. Likewise, both strategies led to a decrease in average solve time, and combining both strategies led to the lowest average solve time.

\begin{table}[h!]
    \centering
    \caption{Performance Comparison on 1000 KSAT Benchmarks}
    \label{tab:table}
    \begin{tabular}{lrr}
        \toprule
        \textbf{Configuration} & \textbf{Solved Instances} & \textbf{Average Solve Time (seconds)} \\ 
        \midrule
        Baseline & 503 & 650.37 \\ 
        DPO only & 612 & 557.63 \\ 
        CDB only & 797 & 376.68 \\ 
        DPO + CDB & \textbf{864} & \textbf{290.56} \\ 
        \bottomrule
    \end{tabular}
\end{table}

\begin{figure}[h!]
    \centering
    \includegraphics[width=\columnwidth]{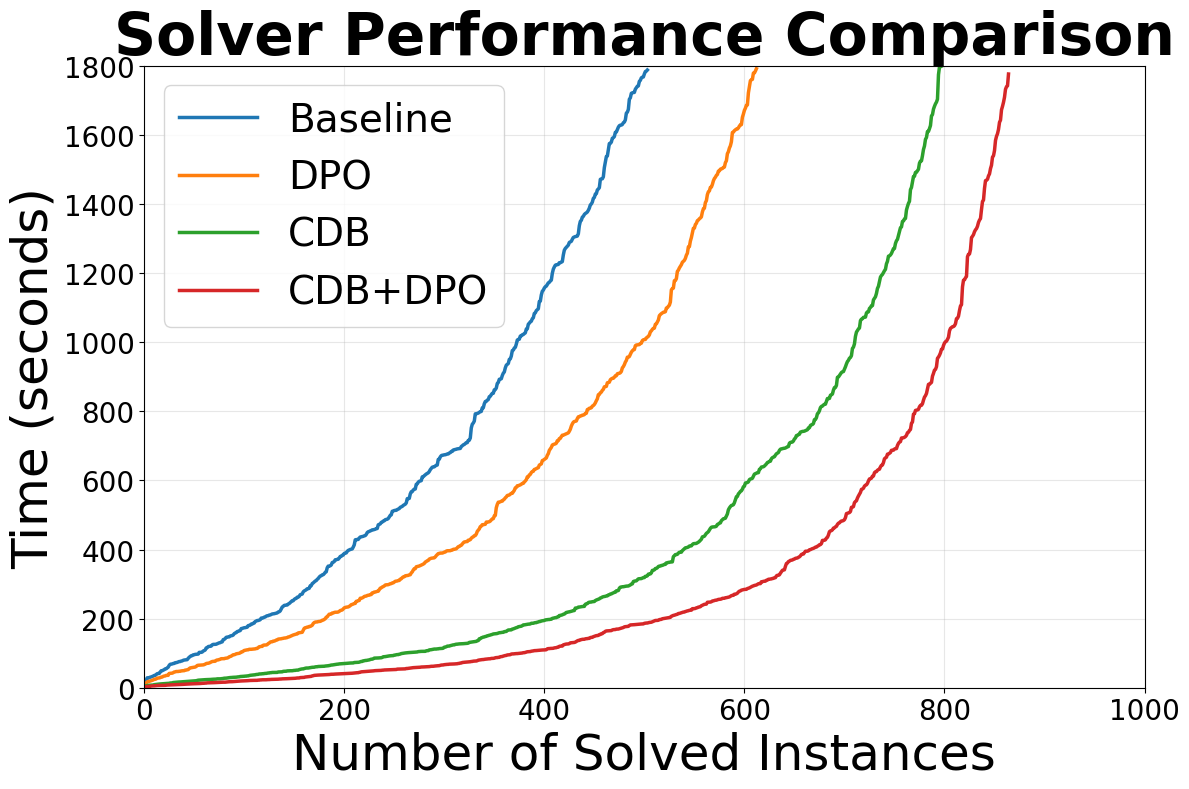}
    \caption{Solver Performance Comparison}
    \label{fig:cactus}
\end{figure}

Figure \ref{fig:cactus} provides cactus plots for a visual comparison of the different strategies. Both DPO and CDB lead to an increase in solved instances in less time, with combining the two strategies yielding the best results.

\section{Conclusion}
In this paper, we introduced two strategies to improve the performance of the St\aa lmarck Procedure. Our results show that the two strategies can significantly improve the St\aa lmarck Procedure. In the future, we plan to further refine the heuristics for branching beyond the simple frequency counting for CDB. This might include learning-based heuristics or lookahead heuristics. Likewise, we plan to explore making DPO dynamic by continually re-evaluating deductive potential instead of using static pre-scoring. As there has been little work done to improve the St\aa lmarck Procedure, there is a wide range of opportunities to improve its efficiency and applicability.

\section{Acknowledgements}
We'd like to extend our gratitude to Professor Haoze Wu and Professor Clark Barrett for their valuable guidance and support.

\bibliographystyle{IEEEtran}
\bibliography{bibli} 

\end{document}